\documentclass[showpacs,showkeys,prb]{revtex4}
\usepackage{amssymb}
\usepackage{amsfonts}
\usepackage{amsmath}
\usepackage{epsfig}
\usepackage{graphicx}

\providecommand{\U}[1]{\protect\rule{.1in}{.1in}}

\begin{document}

\title{Trapping effect of periodic structures on the thermodynamic \\ properties  of Fermi and Bose gases}
\author{P. Salas and M. A. Sol\'is}
\affiliation{Instituto de F\'{\i}sica, UNAM, Apdo. Postal 20-364, 01000 M\'exico D.F.,
M\'exico}


\begin{abstract}

We report the thermodynamic properties of Bose and Fermi ideal gases  
immersed in periodic structures such as penetrable multilayers or multitubes 
simulated by one (planes) or two perpendicular (tubes) external Dirac comb potentials, while the particles are allowed to move freely in the remaining directions. 
Although the bosonic chemical potential is a constant for $T < T_c$, 
a non decreasing  with temperature anomalous behavior of the fermionic chemical 
potential is confirmed and monitored as the tube bundle goes from 2D to 1D when 
the wall impenetrability overcomes a critical value.
In the specific heat curves dimensional crossovers are very noticeable at high temperatures for both gases, 
where the system behavior goes from 3D to 2D and latter to 1D as the 
wall impenetrability is increased. 
\end{abstract}

\maketitle

\section{Introduction}

Non-relativistic quantum fluids (fermions or bosons) constrained by periodic 
structures, such as layered or tubular, are found in many real or man-made physical systems. For example, we find electrons in layered structures such as cuprate high temperature superconductors or semiconductor superlattices, or in tubular structures like organo-metalic superconductors.

On the experimental side, there are a lot of experiments around  
bosonic gases in low dimensions, such as: 
BEC in 2D hydrogen atoms \cite{Safonov1998}, 
2D bosonic clouds of rubidium \cite{Hadzibabic2006}, 
superfluidity in 2D $^4$He films \cite{Bishop1978}, 
while for in 1D we have the confinement of sodium  
\cite{Gorlitz2001}, 
to mention a few. 

 
Meanwhile, for non-interacting fermions there are only a few experiments, 
for example, interferometry probes which have led to observe Bloch 
oscillations 
\cite{Roati2004}. 

To describe the behavior of fermion and boson gases inside this symmetries, several works have been published. For a review of a boson gas in optical lattices see \cite{Dalfovo1999}, 
and for fermions \cite{Giorgini2008} is very complete. 
Most of this theoretical works use 
parabolic \cite{Bagnato1991,Petrov2000}, 
sinusoidal \cite{WK,HR} and  biparabolic \cite{Muradyan} 
potentials, with good results only in the low particle energy limit, 
where the tight-binding approximation is valid. 

Although in most of the articles mentioned above the interactions between particles and the periodic constrictions are taken simultaneously in the system description, the complexity of the many-body problem leads to only an approximate solution. So that the effects of interactions and constrictions in the properties of the system, are mixed and indistinguishable.

In this work we are interested in analyzing the effect of the structure on the properties of the quantum gases regardless of the effect of the interactions between the elements of the gas,  
which we do as precisely as the accuracy of the machines allows us to do.

This paper unfolds as follows: in Sec. 2 we describe our model which consists of quantum particles gas in an infinitely large box where we introduce layers of null width separated by intervals of periodicity $a$. 
In Sec. 3 we obtain the 
grand potential for a boson and for a fermion gas either inside a multilayer or a multitube structure. From these grand potentials we calculate the chemical 
potential and specific heat, which are compared with the 
properties of the infinite ideal gas. In Sec. 4 we discuss results,
and give our conclusions.

\section{Quantum gases whithin multilayers and multitubes}

We consider a system of $N$ non-interacting particles, either fermions 
or bosons, with mass $m_b$ for bosons or $m_f$ for fermions respectively, 
within layers or tubes of separation $a_i$, $i$ = $x$ or $y$, and width 
$b$, which we model as periodic arrays of delta potentials either in the 
$z$-direction and free in the other two directions for planes, and two 
perpendicular delta potentials in the $x$ and $y$ directions and free 
in the $z$ one for tubes. 
The procedure used here is described in detail in Refs. 
\cite{PatyJLTP2010,Paty2} and \cite{SalasJLTP12} for a boson gas, where 
we model walls in all the constrained directions using \textquotedblleft 
Dirac comb\textquotedblright\ potentials. 
In every case, the Schr\"{o}dinger equation for the  
particles is separable in $x$, $y$ and $z$ so that 
the single-particle energy as a function of the momentum $\mathbf{k} = 
(k_{x},k_{y},k_{z})$ is $\varepsilon _{k} = \varepsilon _{k_{x}} + 
\varepsilon _{k_{y}} + \varepsilon_{k_{z}}$. For the directions where 
the particles move freely we have the customary 
dispertion relation $\varepsilon _{k_{i}}=\hbar ^{2}k_{i}^{2}/ 2m_i$,
with $k_{i}=2\pi n_{i}/L$, $n_{i}=\pm 1,\pm 2,...$, and we are assuming 
periodic boundary conditions in a box of size $L$. 
Meanwhile, in the constrained directions, $z$ for planes and $x,y$ for tubes, 
the energies are implicitly obtained through the transcendental 
equation \cite{KP}
\begin{equation}
(P_{i}/\alpha _{i}a_{i})\sin (\alpha _{i}a_{i})+\cos (\alpha
_{i}a_{i})=\cos (k_{i}a_{i}), \label{KPSol}
\end{equation}
with $\alpha_{i}^{2}=2m_i\varepsilon _{i}/\hbar ^{2}$, and the dimensionless 
parameter $P_{i}=m_iV_{0i}a_{i}/\hbar ^{2}$ represents the layer impenetrability in terms of the strength of the delta potential $V_{0i}$. 
We redefine $P_{i}=(m_iV_{0i}\lambda _{F0}/\hbar ^{2})(a_{i}/\lambda _{F0}) 
\equiv P_{0i}(a_{i}/\lambda _{F0})$, where $\lambda _{F0}\equiv 
h/\sqrt{2\pi m_ik_{B}T_{F0}}$ is the thermal wave length of an ideal gas 
inside an infinite box, with $ k_B T_{F0}=E_{F0} = (3\pi^{2})^{2/3}
(\hbar^2/2m_i) \rho ^{2/3} $ the Fermi energy and $\rho =  (k_B T_{F0})^{3/2} 
/3\pi^2 a^2  (\hbar^2/2m a^2)^{3/2}$ is the density of the gas. 

The energy solution of Eq. (\ref{KPSol}) for 
has been extensively analized in Refs. \cite{PatyJLTP2010,Paty2} and 
\cite{SalasJLTP12}, 
where the allowed and forbidden energy-band structure is shown, and the 
importance of taking the full band spectrum has been demonstrated.

\section{Thermodynamic properties of quantum gases in multilayers and 
in multitubes}

Every thermodynamic property may be obtained starting from the grand 
potential of the system under study, whose generalized form is \cite{Path} 
\begin{equation}
\Omega (T,L^{3},\mu )= U - TS - \mu N = \delta_{a,-1} \Omega_0
-\frac{k_{B}T}{a} \sum_{\mathbf{k}{= 0}}
\ln \bigl\{1 + a\exp [-\beta
(\varepsilon _{\mathbf{k}}-\mu )]\bigr\},  \label{omega}
\end{equation}
where $a = -1$ for bosons, 1 for fermions and 0 for the classical 
gas, $\delta$ is the Kronecker delta function and $\beta = 1/k_{B}T$. 
The ground state contribution $\Omega_0$, which is representative 
of the Bose gas, is not present when we analyze the Fermi gas.

For a boson gas inside multilayers we go through the algebra described 
in \cite{PatyJLTP2010,Paty2}, and 
taking the thermodynamic limit one arrives to 
\begin{eqnarray}
\Omega \left( T,L^3,\mu \right) &= &k_{B}T\ln \bigl(1-
\exp [-\beta (\varepsilon _{0}-\mu )]\bigr) \nonumber \\
&& -\frac{1}{\beta ^{2}}\frac{L^{3}m}{\left( 2\pi \right)^{2}\hbar ^{2}} 
{\int_{-\infty }^{\infty }dk_{z}}g_{2}
\bigl\{\exp [-\beta (\varepsilon _{k_{z}}-\mu)]\bigr\}.  
\label{omegaboson}
\end{eqnarray}
Meanwhile, for a fermion gas we get
\begin{equation}
\Omega \left( T,L^{3},\mu \right) =-2\frac{L^{3}}{\left( 2\pi \right) ^{2}}
\frac{m}{\hbar ^{2}}\frac{1}{\beta ^{2}}{\int_{-\infty }^{\infty }dk_{z}} \ 
\mathsf{f}_{2}\bigl\{\exp [-\beta (\varepsilon _{k_{z}}-\mu )]\bigr\},  
\label{omegafermion}
\end{equation}
where $\mathsf{g}_{\sigma}(t)$ and $\mathsf{f}_{\sigma }(t)$ are the Bose 
and Fermi-Dirac functions \cite{Path}. The spin degeneracy has been taken 
into account for the development of Eq. (\ref{omegafermion}).

On the other hand, for a multitube structure we have
\begin{eqnarray}
\Omega \left( T, L^3,\mu \right)  &=& k_{B}T\ln [1-e^{-\beta (\varepsilon
_{0}-\mu )}] \nonumber \\
&& -\frac{L^{3}m^{1/2}}{\left( 2\pi \right) ^{5/2}\hbar }
\frac{1}{\beta ^{3/2}}
\int_{-\infty }^{\infty }\int_{-\infty }^{\infty }dk_{x}\
dk_{y}g_{3/2}(e^{-\beta (\varepsilon _{k_{x}}+\varepsilon _{k_{y}}-\mu )})
\label{Tubosboson}
\end{eqnarray}
for a boson gas, and 
\begin{equation}
\Omega \left( T, L^3,\mu \right) = -2 \frac{L^{3}m^{1/2}}{\left( 2\pi \right)
^{5/2}\hbar }\frac{1}{\beta ^{3/2}}   
\int_{-\infty }^{\infty }\int_{-\infty }^{\infty }dk_{x}\
dk_{y}\mathsf{f}_{3/2}\bigl\{\exp[-\beta (\varepsilon _{k_{x}}+\varepsilon _{k_{y}}-\mu )]\bigr\}
\label{Tubosfermion}
\end{equation}
for a fermion gas.

For calculation matters, it is useful to split the infinite integrals 
into an number $J$ of integrals running over the energy bands, 
taking $J$ as large as necessary to acquire convergence.

\subsection{\bf Chemical potential and specific heat}

\begin{figure}[h]
 \begin{center}
  \includegraphics[scale=0.4]{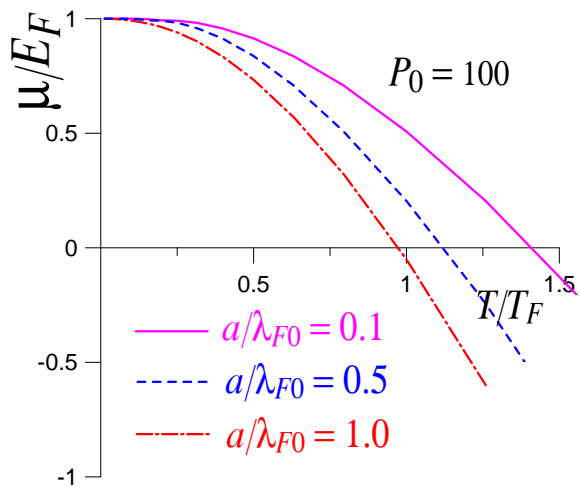}
   \caption{ (Color online) Chemical potential as a function of  $T/T_{F0}$ for planes with $P_0=100$ and different values of $a/\lambda_{F0}$.}
  \end{center}
\label{muvsT}
\end{figure}

For a gas inside a multilayer structure, the particle number 
$N$ is directly obtained from Eqs. (\ref{omegaboson}) and 
(\ref{omegafermion}). Important characteristics can be 
extracted, such as the critical temperature for a condensating boson gas 
and the influence of the system parameters on it, $a$ and 
$P_0$, already reported in  Refs. \cite{PatyJLTP2010,Paty2}. But for 
the case of a fermion gas, we focus on the chemical potential 
since it is closely related to the Fermi energy of the system. In this 
case the number equation is 
\begin{equation}
N= 2\frac{L^{3}}{\left( 2\pi \right) ^{2}}\frac{m}{\hbar ^{2}}\frac{1 }
{\beta ^{2}}{\int_{-\infty }^{\infty }dk_{z}}\ln \bigl\{1+\exp [-\beta
(\varepsilon _{k_{z}}-\mu )]\ \bigr\}, 
\label{NumFer}
\end{equation}
from which we are able to numerically extract the Fermi energy of the 
system, which corresponds to the chemical potential 
for $ T= 0$, over the Fermi energy of the IFG $E_{F0} = (\hbar^2/2m) 
k_{F0}^{2} = (\hbar^2/2m) 4\pi^2/ \lambda_{F0}$, namely $E_F/E_{F0}$, 
as a function of the impenetrability parameter $P_0$, whose behavior 
corresponds to a monotonically increasing curve as $P_0$ increases, 
being more evident for smaller values of $a/\lambda_{F0}$. 
Another important feature is the chemical potential of the system over 
its Fermi energy, $\mu/E_{F}$ as a 
function of the temperature in units of the  Fermi temperature $T/T_{F}$, 
Fig \ref{muvsT}, for a given value of $P_{0}$. 
There is a special interest in Fig \ref{muvsT}, since for certain 
geometrical configurations the chemical potential shows an anomalous 
behavior, as will be shown later. Also, in this last figure one may notice 
that for $P_0 = 0$ the 3D IFG behavior for the chemical 
potential is recovered, and that the curve crosses the $x$ axis in 
$T/T_F = 0.989$, as has been reported in  \cite{Grether}. Meanwhile, as 
$P_0 \rightarrow \infty$, we have a fermion gas inside a two dimension structure, giving a zero chemical potential at $T/T_F = 1.44$, 
as expected.


\begin{figure}[h]
  \begin{center}
   \includegraphics[scale=0.4]{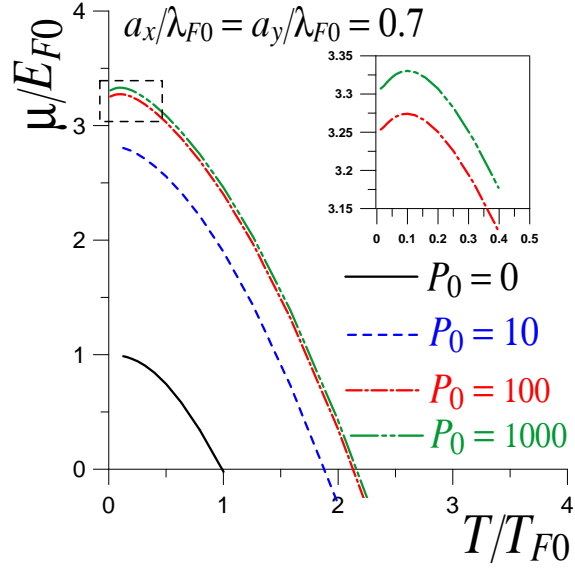}
    \caption{ (Color online) Chemical potential as a function of $T/T_{F0}$ 
for multitubes with $a_{x}/\lambda_{F0} = a_{y}/\lambda_{F0} = 0.7$ and different values of $P_0$.} 
\label{muvsTtubos}
  \end{center}
\end{figure}


We make a similar procedure for the boson an fermi gases inside a 
multitube structure, the first one being reported in \cite{SalasJLTP12}. 
But for a fermion gas we start from the equation 
\begin{equation}
N=2\frac{L^{3}}{\left( 2\pi \right) ^{5/2}}\frac{m^{1/2}}
{\hbar }\frac{1}{\beta^{1/2} }{\int_{-\infty }^{\infty }dk_{z}}
\mathsf{f}_{1/2}\bigl\{\exp [-\beta (\varepsilon_{k_{x}}+
\varepsilon _{k_{y}}-\mu )]\bigr\}  \label{Fermtubos}
\end{equation}
and extract the chemical potential over the Fermi energy of the system, 
$\mu/E_{F}$, which is probably the feature that attracts greater attention 
due to its anomalous behavior shown in Fig. \ref{muvsTtubos}, which shows 
up as an unexpected small hump.

Another interesting characteristic is that the chemical potential over 
the IFG Fermi energy in every case 
is lifted as $P_0$ increases due to the presence of the layers, in the same 
way as the chemical potential of the boson gas started above zero.


The {\it specific heat} of a boson gas has been reported in Ref. \cite{PatyJLTP2010,Paty2,SalasJLTP12}
where we can observe a transition from a 3D system to a 2D one, 
which becomes evident for certain parameter values and sufficiently 
high temperatures. At this point is where the advantages of summing 
over a great amount of allowed energy bands shows its relevance.

Meanwhile, the specific heat for a fermion gas inside layered arrays is 
obtained going through the derivatives of Eqs. (\ref{omegaboson}) 
and (\ref{Tubosfermion}), leading, after some algebra, to 
\begin{eqnarray}
\frac{C_{V}}{Nk_{B}} &=&  \frac{L^{3}}{N\left( 2\pi \right) ^{2}}\frac{m}
{\hbar ^{2}}\bigl\{ \frac{4}{\beta }{\int_{-\infty }^{\infty }dk_{z}}\mathsf{f}_{2}
\bigl\{\exp[-\beta (\varepsilon _{k_{z}}-\mu )]\bigr\}  \nonumber \\
&&+2{\int_{-\infty }^{\infty }dk_{z}} 
\ln \bigl\{1+\exp [-\beta (\varepsilon
_{k_{z}}-\mu )]\bigr\} \{2\varepsilon _{k_{z}}-\mu +T\frac{d\mu }{dT}\}  \nonumber \\
&&+2\beta {\int_{-\infty }^{\infty }dk_{z}}\frac{\varepsilon _{k_{z}}
\bigl\{\varepsilon_{k_{z}}-\mu +T\frac{d\mu }{dT}\bigr\}}{\exp 
\bigl\{\beta (\varepsilon_{k_{z}}-\mu )\bigr\}+1} \bigr\} 
\label{Cvfermiones}
\end{eqnarray}
for multiplanes, and 
\begin{eqnarray}
\frac{C_{V}}{Nk_{B}}&=&\frac{2 L^{3}}{N\left( 2\pi \right)^{5/2}}
\frac{m^{1/2}}{\hbar} \bigl\{ \frac{\beta ^{1/2}}{2}\int_{-\infty}^{\infty }
\int_{-\infty}^{\infty }dk_{x}dk_{y}\times f_{1/2}(e^{-\beta 
(\varepsilon _{k_{x}}+\varepsilon _{k_{y}}-\mu)})  \nonumber \\
&&(2\varepsilon _{k_{x}}+2\varepsilon _{k_{y}}-\mu +T\frac{d\mu }{dT}) 
+ \beta^{3/2}\int_{-\infty}^{\infty }\int_{-\infty}^{\infty }dk_{x} dk_{y}
(\varepsilon _{k_{x}}+\varepsilon _{k_{y}})\times  \nonumber \\
&&f_{-1/2}(e^{-\beta (\varepsilon _{k_{x}}+\varepsilon _{k_{y}}-\mu)})
\{\varepsilon _{k_{x}}+\varepsilon _{k_{y}}-\mu +T\frac{d\mu }{dT}\} \nonumber \\
&&+ \frac{3}{4 \beta ^{1/2}} \int_{-\infty}^{\infty }
\int_{-\infty}^{\infty }dk_{x}\ dk_{y}f_{3/2}(e^{-\beta
(\varepsilon _{k_{x}}+\varepsilon _{k_{y}}-\mu )}) \bigr\}
\end{eqnarray}
for multitubes.

\begin{figure}
  \begin{center}
       \includegraphics[scale=0.4]{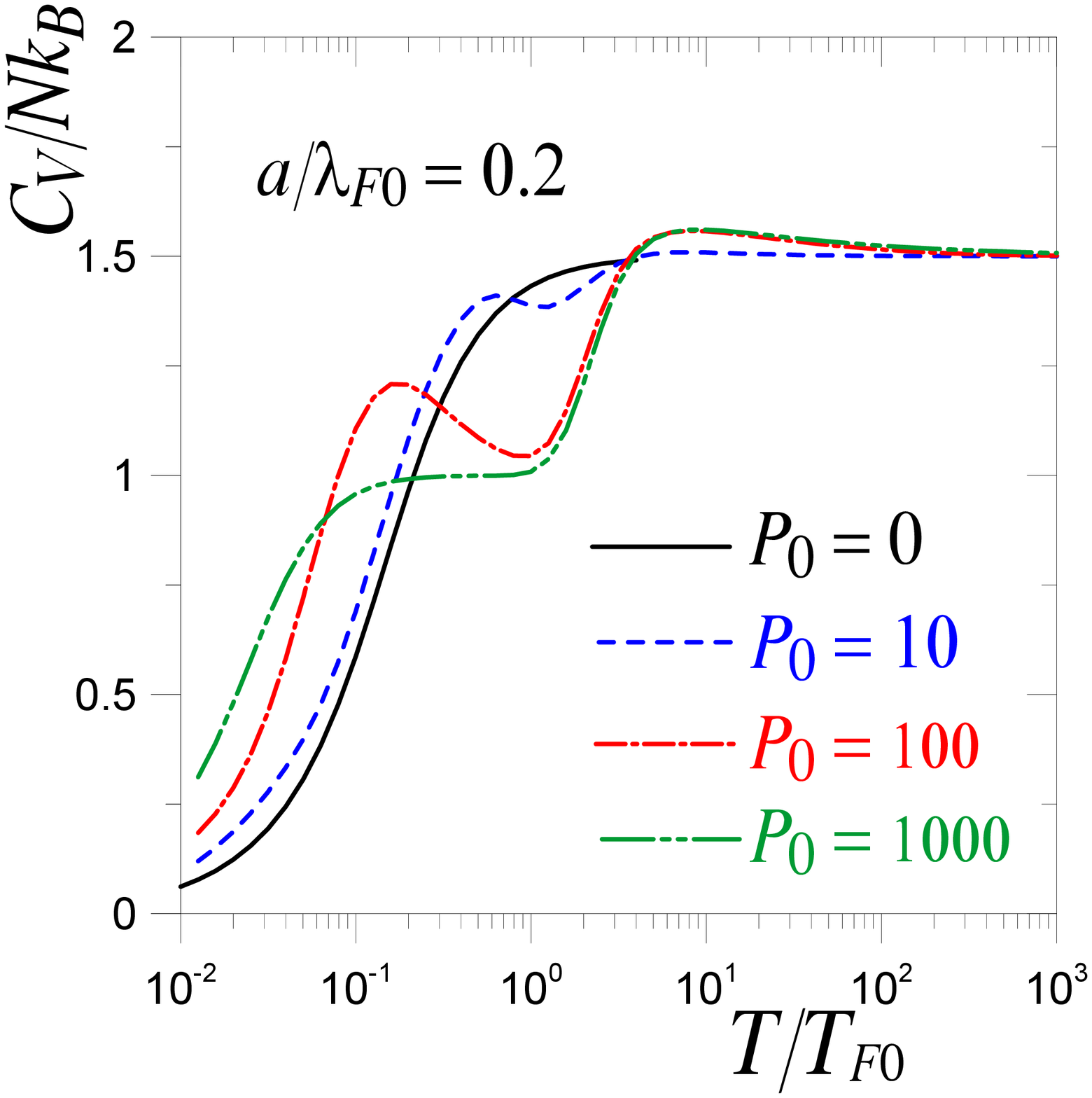}
   \caption{(Color online) Specific heat as a function of $T/T_{F0}$ 
for a fermion gas in multilayers.}
\label{Cvlayers}
  \end{center}
\end{figure}

\begin{figure}
  \begin{center}
  \includegraphics[scale=0.4]{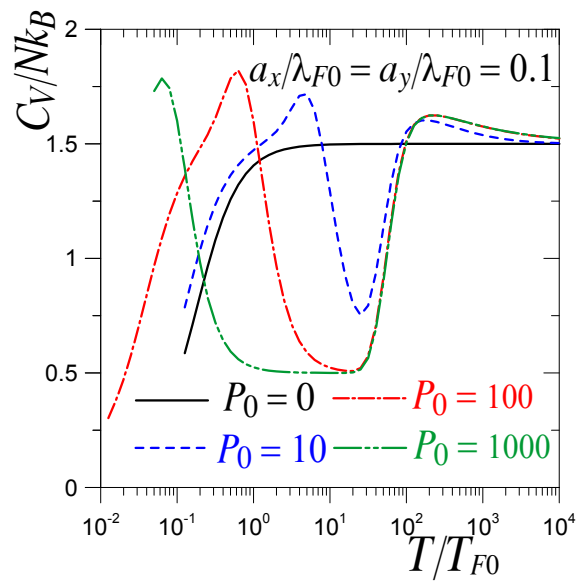}
    \caption{(Color online) Specific heat as a function of $T/T_{F0}$ 
for a fermion gas in multitubes.} 
\label{Cvtubos}
  \end{center}
\end{figure}

In Figs. \ref{Cvlayers} and \ref{Cvtubos} we show the 
behavior of the specific heat of layers with a separation among layers of 
$a/\lambda _{0}=0.2$ and several 
impenetrability intensities $P_0$, as a function of the temperature over 
the system's Fermi temperature. It may be observed that, as the barriers 
disappear ($P_0 = 0$), one recovers the IFG specific heat classical value, 
$3/2$. It is also noticeable that as the value of $a/\lambda _{0}$ 
diminishes, a dimensional crossover signature from 3D to 2D becomes evident, since the the first minimum (going from right to left) in the 
specific heat deepens towards a value $C_V/Nk_B = 1.0$ and broadens as 
$P_0$ increases.
In Fig. \ref{Cvtubos} the dimensional crossover from 3D to 1D is very 
noticeable as the mentioned minimum drops down to the value $1/2$ which 
corresponds to one-dimension.

\vspace{-0.5cm}

\section{Conclusions}

In summary, we have calculated the thermodynamic properties of ideal
boson and fermion gases inside periodical structures. In our model the
multilayers and multitubes are generated with Dirac-comb potentials in 
either one or two directions, while the particles are free in the 
remaining directions.
Just by introducing the planes, 
the translational symmetry of the particles is broken. This fact reflects in every thermodynamic property of the constrained system. In particular, fermions in multi-tubes progress from a 3D behavior to that 2D and finally to 1D as the wall impenetrability is increased, which is observed in the curves of the specific heat as a function of temperature. There is a critical value of the wall impenetrability for which the system begins to behave in dimensions less than two, which is signaled by the appearance of an anomalous chemical 
potential. Bosons in multitubes show similar dimensional crossover like 
that expressed by fermions, in addition to the Bose-Einstein condensation at temperatures below the critical temperature of a ideal Bose-gas with the same particle density. \\   



\noindent {\bf Acknowledgements.}
We acknowledge partial support from UNAM-DGAPA-PAPIIT (M\'{e}xico) \#
IN105011 and IN111613-3, and CONACyT 104917.

\end{document}